 \documentclass[final,5p,times,twocolumn,number,compress]{elsarticle}

\usepackage{amssymb}
\usepackage{amsmath}
\PassOptionsToPackage{hyphens}{url}
\usepackage[usenames,dvipsnames]{color}
\usepackage{setspace}
\usepackage{caption}
\usepackage{subcaption}
\usepackage{ulem}
\usepackage[T1]{fontenc}

\usepackage{hyperref}
\hypersetup{
    citecolor=blue,
    colorlinks=true,
    linkcolor=blue,
    filecolor=blue,      
    urlcolor=black,
    pdfpagemode=FullScreen,
}

\journal{Physics Letters B}

\makeatletter
\def\ps@pprintTitle{%
\def\@oddhead{\hfill Preprint number: TTP23-044, DESY-23-149}%
 \let\@evenhead\@empty
 \let\@evenfoot\@oddfoot}

\gdef\emailauthor#1#2{\stepcounter{ead}%
     \g@addto@macro\@elseads{\raggedright%
      \let\corref\@gobble\def\@@tmp{#1}%
      \eadsep{\ttfamily\expandafter\strip@prefix\meaning\@@tmp}
      \def\eadsep{\unskip,\space}}%
}

\makeatother

\begin{document}

\begin{frontmatter}



\title{Riding the dark matter wave: Novel limits on general dark photons from LISA Pathfinder
}

\author[desy]{Jonas Frerick\corref{cor1}}
 \ead{jonas.frerick@desy.de}
\affiliation[desy]{%
organization={ Deutsches Elektronen-Synchrotron DESY}, addressline={ Notkestr. 85}, postcode={22607}, city={Hamburg}, country={Germany}
}%

\author[heidelberg]{Joerg Jaeckel}
\ead{jjaeckel@thphys.uni-heidelberg.de}
\affiliation[heidelberg]{organization={Institut f\"{u}r Theoretische Physik, Universit\"{a}t Heidelberg}, addressline={
Philosophenweg 16}, postcode={69120} ,city={Heidelberg}, country={Germany}
}%

\author[kit]{Felix Kahlhoefer}
\ead{kahlhoefer@kit.edu}
\affiliation[kit]{organization={Institute for Theoretical Particle Physics (TTP),
Karlsruhe Institute of Technology (KIT)}, postcode={ 76128},city ={Karlsruhe}, country={Germany}
}%
\author[desy]{Kai Schmidt-Hoberg}
\ead{kai.schmidt-hoberg@desy.de}
\cortext[cor1]{Corresponding author}

\begin{abstract}
We note the possibility to perform a parametrically improved search for gauged baryon ($B$) and baryon minus lepton ($B-L$) Dark Photon Dark Matter (DPDM) using auxiliary channel data from LISA Pathfinder. 
In particular we use the measurement of the differential movement between the test masses (TMs) and the space craft (SC) which is nearly as sensitive as the tracking between the two TMs. TMs and SC are made from different materials and therefore have different charge-to-mass ratios for both $B-L$ and $B$. Thus, the surrounding DPDM field induces a relative acceleration of nearly constant frequency.
For the case of $B-L$, we find that LISA Pathfinder can constrain previously unexplored parameter space, 
providing the world leading limits in the mass range $4\cdot 10^{-19}\,\text{eV}<m<3\cdot 10^{-17}\,\text{eV}$. This limit can easily be recast also for dark photons that arise from gauging other global symmetries of the SM.
\end{abstract}


\end{frontmatter}

\section{Introduction}\label{sec:intro}

The existence of Dark Matter (DM) is a well-established observational fact \cite{Bertone:2016nfn}. For a long time the WIMP paradigm has dominated the quest for DM \cite{Lin:2019uvt} but with null observations in the increasingly sensitive direct detection (DD) experiments \cite{XENON:2023cxc,LZ:2022lsv,Liu:2022zgu} there is increased interest in alternative DM candidates. One particularly well-motivated class of such particles are ultra-light and weakly coupled bosons (see, e.g.,~\cite{Jaeckel:2010ni,Ringwald:2013via,Dobrich:2015xca} for reviews). These include, among others, the axion \cite{Peccei:1977hh,Weinberg:1977ma,Wilczek:1977pj,Marsh:2015xka} or general axion-like particles (ALPs) \cite{Masso:2002ip} as well as new vector bosons (cf., e.g.,~\cite{Fayet:1980rr,Fayet:1990wx, Jaeckel:2012mjv,Dror:2017ehi,Bauer:2018onh,Fabbrichesi:2020wbt}), often referred to as dark photons (DPs).\footnote{In this work we will refer to any light new vector boson as a dark photon, allowing for couplings that gauge global symmetry groups of the SM. Our method is unfortunately insensitive to the ``canonical'', kinetically mixed DP.}

In this work, we will focus on ultra-light DPs as a DM candidate. Small DP masses $m$ can be generated either by the St\"{u}ckelberg \cite{Stueckelberg1938,Ruegg:2003ps} or by the Higgs mechanism (where the latter often causes additional constraints). For $m \lesssim 30\,$eV the local DM halo behaves like a classical wave as the spacing between particles becomes smaller than the de Broglie wave length \cite{Hui:2021tkt}. We will stay agnostic about the details of the production mechanism. While heavier DPs can be produced from the thermal SM bath \cite{Redondo:2008ec}, very light DPs require a non-thermal mechanism to ensure that the DM is cold. To this end, there are many gravitational or extended dark sector solutions that provide the correct relic density \cite{Nelson:2011sf,Arias:2012az,Graham:2015rva,Bastero-Gil:2018uel,Agrawal:2018vin,Co:2018lka,Dror:2018pdh,Long:2019lwl,Nakayama:2019rhg,Nakai:2020cfw,Ahmed:2020fhc,Kolb:2020fwh,Salehian:2020asa,Firouzjahi:2020whk}.  Furthermore, we will assume that the DP is gauged under a combination of baryon number ($B$) and lepton number ($L$), with a particular focus on the difference ($B-L$).

Astrophysical objects like the Sun can efficiently produce light DPs which, in turn, results in impressive limits on the existence of DPs without requiring them to be DM \cite{Redondo:2008aa,An:2013yfc,Hardy:2016kme,Li:2023vpv,Frerick:2022mjg}. Additionally, lab experiments testing the equivalence principle are perfect candidates to look for this kind of new physics that induces long-range forces beyond electromagnetism and gravity \cite{Wagner:2012ui,Fayet:2017pdp,Fayet:2018cjy,MICROSCOPE:2022doy}. Planetary \cite{KumarPoddar:2020kdz} and asteroidal \cite{Tsai:2021irw,Tsai:2023zza} orbits are sensitive to new long-range forces as well. Finally, the gauge anomaly associated with baryon number leads to strong constraints from meson decays \cite{Dror:2017ehi} for this specific gauge group.

If the DPs are also DM, new tests become available, e.g.\ with accelerometers as proposed in Ref.~\cite{Graham:2015ifn} and realized in~\cite{Shaw:2021gnp}. A particularly interesting possibility is to search for these DM candidates directly at gravitational wave observatories. This idea was first pointed out in Ref.~\cite{Arvanitaki:2014faa} for ultra-light scalar DM and later briefly discussed for DPDM in Ref.~\cite{Graham:2015ifn}. Independently, Ref.~\cite{Pierce:2018xmy} focused especially on GW interferometry and performed a more detailed analysis for several instruments based on the small inhomogeneity of the field. We will use this work as a guideline for our own analysis as it also investigated new vector bosons gauged under $B$ and $B-L$, and discussed both ground-based and space-based laser interferometers. 
However, it makes use of the differential acceleration between the two equal test masses and is therefore limited by the small ratio of arm length to the scale of inhomogeneity. Here, we point out that the setup of LISA Pathfinder (LPF) also offers the possibility to use the differential acceleration between a test mass and the satellite carrying the interferometer itself. As we will argue this type of search is not limited by the arm length, significantly increasing the sensitivity in the relevant mass range compared to a previous analysis~\cite{Miller:2023kkd}. That said, we want to point out that the use of auxiliary channels was already proposed for KAGRA in Ref.~\cite{Michimura:2020vxn}.
 
We briefly discuss the signal prediction and analysis method in sec.~\ref{sec:signal} and point out the similarities and differences to previous DPDM interferometer limits. This is followed by an introduction to LPF including a discussion of the sensitivity in sec.~\ref{sec:LPF}. Finally, we estimate the sensitivity of this instrument to DPDM and discuss the necessary steps for a refined analysis in sec.~\ref{sec:res} before we conclude in sec.~\ref{sec:con}. Throughout this letter we work in natural units $\hbar=c=1$.

\section{Calculation of the signal}\label{sec:signal}

Let us begin by introducing the DP Lagrangian \cite{Holdom:1985ag,Babu:1997st,Bauer:2018onh}

\begin{equation}
    \mathcal{L}\supset-\frac{1}{4}F_{\mu\nu}^{\prime}F^{\prime\mu\nu}-\frac{\epsilon_\text{KM}}{2}F_{\mu\nu}^{\prime}F^{\mu\nu} + \frac{m^2}{2}A_\mu^\prime A^{\prime \mu} -\epsilon_g e A_\mu^\prime J^\mu_g\;,
\end{equation}
which contains the renormalisable interactions of the DP field $A^\prime_\mu$ in full generality. It takes into account both the kinetic mixing $\epsilon_\text{KM}$ between the field strength tensors $F_{\mu\nu}^{(\prime)}$ of the SM photon and the dark photon, and an explicit coupling $\epsilon_g$ to a current $J^\mu_g$ associated with a gauge group $g$. Note that we have rescaled the gauge coupling $g_g$ to the electromagnetic coupling $e$, i.e. $g_g=\epsilon_g e$. From now on, we will assume the kinetic mixing to be negligible and focus on the explicit couplings with $g=B$ or $g=B-L$ unless mentioned otherwise. 
In sec.~\ref{sec:res} we will discuss how to generalize our analysis to arbitrary gauge groups.

Under the aforementioned assumptions, any piece of baryonic matter is directly charged under both gauge groups. Thus, in a background field of DPDM these charges behave in full analogy to electric charges in an electric field. Assuming the DM to be cold and have mass $m$, the field will be nearly monochromatic with a linewidth suppressed by the non-relativistic velocity $v\sim 10^{-3}$ of the halo \cite{Evans:2018bqy}
\begin{equation}
    \mathbf{A}(t,x)=\mathbf{A}_\text{DM}e^{-i\omega t +\phi(x)}\;,\label{eq:3pot}
\end{equation}
where $\omega= m + \mathcal{O}(v^2)$ denotes the non-relativistic particle energy, $\mathbf{A}_\text{DM}$ is the 3-vector of the DPDM field, and $\phi(x)=i\;\mathbf{k}\cdot\mathbf{x}+\phi_0$ is a weakly position dependent phase where $\mathbf{k}\approx m\mathbf{v}$ denotes the momentum of the wave and $\phi_0$ is a constant phase.\footnote{For our purposes, weakly dependent means that $|\mathbf{k}|L\ll 1$ where $L$ denotes the size of the experiment.} We can obtain the temporal component of the 4-potential from the ``Lorenz condition''
\begin{equation}
    \partial_\mu A^\mu=0 \Rightarrow A^0(t)= -\frac{\mathbf{k}\cdot \mathbf{A}(t)}{\omega}\approx  \mathbf{v}\cdot \mathbf{A}(t)\ll |\mathbf{A}(t)|\;,
\end{equation}
which has to be fulfilled for a massive vector boson as dictated by the equations of motion. Due to the weak spatial dependence of eq.~\eqref{eq:3pot} we have omitted the $x$ in the argument of all vector components. We observe that the temporal component is generically velocity suppressed for non-relativistic DPDM.\footnote{Indeed, for transversely polarized DPs, i.e.\ $\mathbf{k}\cdot \mathbf{A}_\text{DM}=0$, the component vanishes exactly.}

Within a coherence patch the signal can be treated as being monochromatic. The coherence length is given by the wavelength $\lambda_c\simeq 2\pi/(mv)$ and the coherence time is $t_c\simeq 2\pi/(mv^2)$. It is important to realize that neither the amplitude nor the direction of the field changes within a coherence patch. LISA Pathfinder (LPF) covers a frequency range from a few Hz down to around $10^{-5}\,$Hz (see sec.~\ref{sec:LPF}). Especially for the lowest frequencies this implies an extremely long coherence time due to the non-relativistic velocities.
In fact, even for the highest frequencies in LPF's sensitivity range the coherence time is more than a week so that even long-term searches for monochromatic signals suffer at most weakly from the decoherence of the signal, thus enhancing the limits significantly without employing new technology.

\bigskip

For the sake of clarity, we will give a minimal description of LPF here and use it as an example for the main idea behind our analysis but the following arguments are equally valid for two generic objects made from different materials. In LPF, two (almost) identical test masses (TMs) are enclosed separately in a space craft (SC) and the relative motion between the TMs themselves and between TMs and the SC is tracked. A more detailed description will follow in sec.~\ref{sec:LPF}. The dominant effect of the gauged DPDM on a charged object is analogous to the electric component of the Lorentz force (see~\cite{Graham:2015ifn} for a similar calculation as we do in the following). Therefore, we need to determine the ``electric field''
\begin{align}
    \mathbf{E}_g=-\partial_t \mathbf{A}(t)=i\omega \mathbf{A}_\text{DM}e^{-i\omega t+\phi(x)}\;.\label{eq:efield}
\end{align}
For our purposes, we can completely ignore the phase of the field $\phi(x)\approx 0$ and consider it spatially constant over the size of the experiment. Therefore, the field is oscillating at a single frequency as long as we consider only coherent time scales.

This field then exerts a force on all objects charged under the given gauge group. Therefore, we find the following acceleration
\begin{align}
    \mathbf{a}(t)\simeq i\omega\epsilon_g e \frac{q}{M}\mathbf{A}_\text{DM}e^{-i\omega t}=i\epsilon_g e\frac{q}{M}\sqrt{2\rho_\text{DM}}\;\hat{\mathbf{e}}_A\; e^{-i\omega t}\;,\label{eq:acc}
\end{align}
for an object of mass $M$ and charge $q$ (under $g$). Furthermore, we used the well-known relation between the average energy density and amplitude of wave-like DM 
\begin{equation}
    \rho_\text{DM}=\frac{1}{2}\omega^2|\mathbf{A}_\text{DM}|^2\;.
\end{equation}
Finally, $\hat{\mathbf{e}}_A$ is the unit vector in direction of the DPDM field within a coherence patch.

To estimate the signal-to-noise ratio in LPF we need the amplitude of the relative acceleration between the SC center of mass and the TMs for the three SC axes $i=x,y,z$. For a monochromatic signal, we obtain those values by taking the real part and dropping the harmonic behavior of eq.~\eqref{eq:acc}
\begin{align}
    \Delta a_i=&\epsilon_g e\left(\Delta \frac{q}{M}\right)\sqrt{2\rho_\text{DM}}\cos\theta_{A,i}\label{eq:axacc}\; .
\end{align}
We note that a one-dimensional setup or even a typical planar interferometer can in principle be totally insensitive to this effect if the polarization is orthogonal to the plane of the experiment. LPF offers the advantage that it features a full 3D sensitivity of the SC motion w.r.t.\ the TMs.

In general, there are two different polarization models in the literature. One of them assumes that the DPs have the same polarization everywhere in space while the other one assumes that each coherence patch has a different polarization which is distributed uniformly on the unit sphere. From our discussion of the coherence time we conclude that in the ultra-low mass regime we will not be able to tell the difference as all experiments with realistic lifetimes will only observe a single coherence patch.\footnote{At the higher end of the frequency range the situation might be more promising if we allow for an observation time of several years. This can get even better if the sensitivity can be extended to higher frequencies.}
In contrast, for larger masses measuring at different times corresponds to measuring different polarization of the DPs. Combining this with the fact that LPF has a non-trivial orbit and orientation will result in a very complex scheme required to perform a rigorous analysis. Nevertheless, applying this information which in principle is known might provide additional constraining power as demonstrated in Ref.~\cite{Caputo:2021eaa}. We will treat this issue in more detail in sec.~\ref{sec:res}. To conclude this discussion we emphasize that the position and orientation of the SC will not change significantly on the time scales of the used observations~\cite{GIULICCHI2013283}.

\bigskip

Let us quickly compare our result, using the auxiliary channels between SC and TMs, to the case where $\Delta \frac{q}{M}=0$ which corresponds to two bodies made from the same material.
To the best of our knowledge, the TMs for the main interferometers in all GW searches including LPF fulfill this criterion. Additionally, any elemental impurities that could break this degeneracy are kept extremely small in order to improve the performance of the interferometer. Therefore, in this case, we have to look for subleading effects, e.g.\ from the phase in eq.~\eqref{eq:efield} which introduces both an arm-length and a velocity suppression. This is exactly the approach following Ref.~\cite{Pierce:2018xmy}. Only later 
it was re-discovered that the finite light-traveling time of the laser \cite{Morisaki:2020gui} leads to an improvement if the length scale associated with the DP mass $1/m$ coincides with the arm length of the interferometer $L$ as already pointed out in Refs.~\cite{Arvanitaki:2014faa,Graham:2015ifn}.\footnote{We thank the anonymous referee for pointing out the historically correct version of how these limits were (re-)derived. At this point, it should be emphasized that this analysis method was already applied directly to LIGO/VIRGO data \cite{Guo:2019ker,LIGOScientific:2021ffg}.} This ``new'' analysis method and the decoherence effect from the small inhomogeneity of the field will give an observable relative acceleration even for strictly equal charge-to-mass ratios. Nevertheless, this acceleration is suppressed by $\max\left\{(\omega L)^2,v\omega L\right\}$. For full scale interferometers where the arm length is on the scale of $1/m\sim 1/\omega$ by construction this may not be big a problem. But, for LPF with its very limited arm length of $\sim 40\,$cm these effects will substantially suppress all limits derived following the standard methods in the literature as shown in Ref.~\cite{Miller:2023kkd}. Therefore, looking for auxiliary channels between TMs and the SC that feature $\Delta \frac{q}{M}\neq 0$ is promising. Indeed, for a different gravitational wave interferometer, KAGRA, this observation was already utilized in Ref.~\cite{Michimura:2020vxn}, which enhanced the limits in the low frequency region significantly.

With the prediction for the acceleration amplitude, it is straightforward to estimate the signal-to-noise ration (SNR) of an interferometer with a given relative acceleration amplitude spectral density (ASD) $S_a^{1/2}(f)$ via
\begin{equation}
    \text{SNR}=\frac{ \Delta a_i}{S_a^{1/2}(f)}\sqrt{T_\text{eff}}\;,\label{eq:snr}
\end{equation}
where $T_\text{eff}$ depends on observation time $T_\text{obs}$ and coherence time $t_c$ via
\begin{align}
    T_\text{eff}=\begin{cases}
        T_\text{obs}\ \ ,& \quad T_\text{obs}\leq t_c\\
        \sqrt{T_\text{obs}t_c}\ \ ,& \quad  T_\text{obs} > t_c
    \end{cases}\; ,\label{eq:teff}
\end{align}
as outlined e.g.\ in the appendix of Ref.~\cite{Budker:2013hfa}.\footnote{Conventionally, results are quoted as ASDs for the relative displacement instead of acceleration as used by us. The translation from our ASD to this so-called strain sensitivity is fairly simple as it just requires a rescaling factor of $\sim \omega^2/L$.}

\section{LISA Pathfinder sensitivity}\label{sec:LPF}

LPF \cite{LISAPathfinder:2017khw} was a precursor mission to the planned space-borne gravitational wave interferometer LISA \cite{LISA:2017pwj}. The mission's objective was to demonstrate that the technology developed for LISA will be able to perform as predicted under realistic space conditions. For this purpose, a SC containing two TMs was sent to the first Lagrange point of the Sun-Earth system. These TMs are $2\,$kg Gold-Platinum alloy cubes with a side length of $\sim 5\,$cm and they were placed in two separate electrode housings with an optical bench placed in between. The main aim was to keep the noise in the relative acceleration between the two free falling TMs at a level that would verify the applicability of this technology for LISA. Indeed, the test was successful and performed even better than expected \cite{Armano:2016bkm,Armano:2018kix}. Such a high-precision instrument requires more scrutiny than just a single interferometer measuring the relative TM displacement. Therefore, LPF contained a radiation monitor \cite{Armano:2017bsn}, additional interferometers \cite{Audley:2017zvj} and capacitive sensing \cite{LISAPathfinder:2017nwk}. Several of these auxiliary channels are used to avoid a collision between ``TM1'', the reference test mass, and the SC. The choice of a preferred TM is required as two TMs on their respective geodesics within a single SC cannot coexist without a collision. Therefore, there are measures in place to correct the trajectory of the second TM w.r.t.\ the reference TM. 

The TMs are aligned on what the collaboration labeled the x-axis. On this axis, there is the so called $x_{12}$ interferometer which is the central instrument on-board as it is used to measure the relative acceleration between the TMs. For our purposes, we want to focus on another instrument, the $x_1$ interferometer, which controls the SC position w.r.t.\ TM1. This auxiliary interferometer will be the best channel to search for DPDM over a large mass/frequency range.

As we have pointed out in sec.~\ref{sec:signal} we require knowledge about the charge-to-mass ratio and therefore the elemental composition of both the TMs and the SC. Unfortunately, the SC itself is made up of a collection of different materials and components but they are on average expected to be at much lower atomic number than Au or Pt and thus they will have different charge-to-mass ratios. This result is intuitive for $B-L$ as the total charge of an atom is given by the neutron number and the neutron-to-proton ratio tends to increase with atomic number yielding different charge-to-mass ratios for light and heavy elements. This effect is more subtle for $B$. The difference mainly comes from the variation in binding energies and the small mass difference between proton and neutron. This immediately provides us with an estimate for the suppression of the charge-to-mass ratio w.r.t.\ the $B-L$ result: both the binding energies and the nucleon mass difference are of order MeV compared to the total nucleon masses which are at the GeV scale. Naively, this suggests a suppression factor $\sim 10^{-3}$ which turns out to be quite accurate, as we will see in Sec.~\ref{sec:res}.

In the composition of the SC, the second most important contribution after the technology package enclosing the TMs arises from the structure of the SC which is made mostly from carbon and aluminium \cite{Racca:2009zz}. Indeed, the SC contains many different sub-components of similar mass and some of them will also contain elements with atomic number much larger than C or Al. A detailed analysis of the SC composition is beyond the scope of this letter and thus we will simply use a lower bound on the charge-to-mass ratio of the SC. To arrive at this conservative estimate, we will assume that all components are made from the same material as the TMs except for the SC structure which we assume to be entirely made from carbon. Using this approximation and table~1 from Ref.~\cite{Racca:2009zz}, we conclude that the $450\,$kg SC has an $83\,$kg C component and the remaining material will have a charge-to-mass ratio equal to that of gold.

\begin{figure}[t]
    \centering
    \includegraphics[scale=1,]{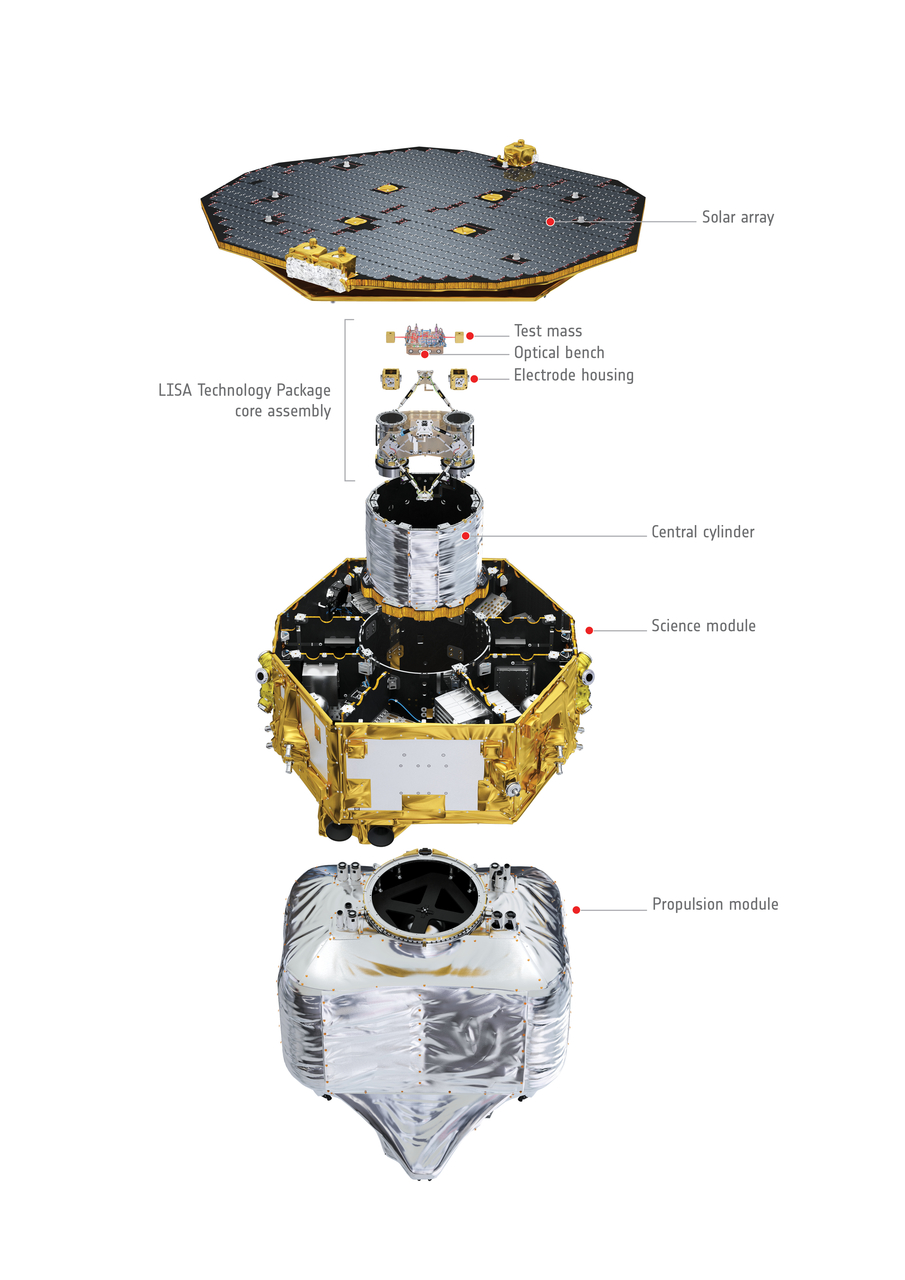}   
    \caption{Exploded view of LPF showing the science module containing the test masses in its center as well as the propulsion module. Image by ESA/ATG medialab (with permission).
    }
    \label{fig:lpf1}
\end{figure}

\begin{figure}[t]
    \centering
    \includegraphics[scale=0.45]{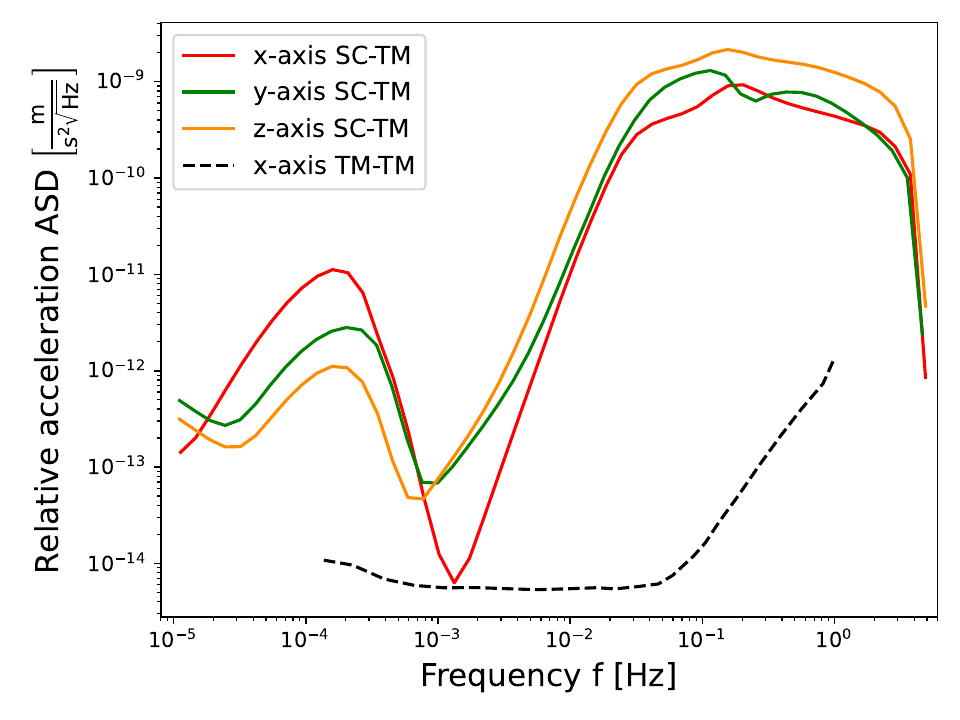}
    \caption{Sensitivity of the LPF SC acceleration w.r.t.\ the TM(s). Red shows the interferometer sensitivity while green and orange are found from the capacitive sensors in the housing averaged over both TMs; data from~\cite{LISAPathfinder:2018sdh}. The dashed black line shows the maximum sensitivity of the LPF TM-TM measurement based on the same data set; taken from~\cite{Armano:2016bkm}}.
    \label{fig:asd}
\end{figure}

 For a better understanding of the geometry we show an exploded view of LPF in fig.~\ref{fig:lpf1}. An important factor in the sensitivity analysis of the SC motion against the TMs is that not just the x-axis but also the y- and z-axis are tracked where the z-axis points from the TMs to the solar array. These axes are measured via capacitive sensing which in general is less precise than the interferometers for most frequencies. Nevertheless, we have the advantage of being able to analyze the relative acceleration ASDs for all SC axes \cite{LISAPathfinder:2018sdh} and we show these results in fig.~\ref{fig:asd}. They represent the simulated, data-backed sensitivities to the relative acceleration of the SC w.r.t.\ the TM(s) which is exactly what we are interested in for eq.~\eqref{eq:snr}.\footnote{In fact, the y- and z-direction is tracked w.r.t.\ the average of both TM coordinates while for the x-axis only the relative motion w.r.t.\ TM1 is measured via the additional interferometer.} These results were obtained from a 6.5 day noise-only run in April 2016 \cite{Armano:2016bkm}. The curves explicitly account for all known noise on SC and TMs and therefore they present the best estimate for the stability of the SC w.r.t.\ the TM(s). To derive limits, we will set cuts at $1\,$Hz and $10^{-4}\,$Hz as a careful evaluation of the highest and lowest frequencies is beyond the scope of this work. Nevertheless, a detailed analysis of the data will most likely lead to interesting constraints in these extremal regimes.

Before we calculate the limits from these ASDs, let us briefly discuss their behavior. For higher frequencies down to $\sim 10^{-3}\,$Hz, the sensitivity is limited mostly by the so-called out-of-loop noise which describes several external influences on the SC. The bump at the low frequency part of the spectrum is due to the star-tracker noise which comes from imperfections in the determination of the position of the SC. Ref.~\cite{LISAPathfinder:2018sdh} argues that this low-frequency noise will most likely be mitigated in the LISA mission pointing out an interesting avenue for future investigations of gauged DPDM. We will discuss this in more detail in sec.~\ref{sec:res}. In the extreme low frequency region we observe additional loss in sensitivity from the capacitive actuation noise experienced by the TMs. In this regime, the simulation also predicts a significantly better sensitivity than the data as shown in Ref.~\cite{LISAPathfinder:2018sdh} further justifying the cuts introduced above. At peak sensitivity, the x-axis almost reaches the TM1-TM2 result, cf.\ the dashed black line in fig.~\ref{fig:asd} taken from Ref.~\cite{Armano:2016bkm} which is based on the same data sample.\footnote{Unfortunately, the full frequency range is not shown in that work.} At this point of best sensitivity the other axes perform comparably worse as the capacitive sensing cannot compete with the interferometer on the x-axis.

\section{Results}\label{sec:res}

\begin{table}[t]
    \centering
    \begin{tabular}{lc c c c c}
    \hline\hline
         Material& & Au & C & SC-TM\\ 
         \hline
         & & & &\\
       $\left(\frac{q}{M}\right)_{B-L}\ \text{in GeV}^{-1}$ & & $0.64$   & $ 0.54 $ &  $ 0.018 $\\
       & & & &\\
       $\left(\frac{q}{M}\right)_{B}\ \text{in GeV}^{-1}$ & &$ 1.0736 $&$ 1.0737 $&$ 1.8\cdot 10^{-5} $\\
       & & & &\\
       \hline\hline
    \end{tabular}
    \caption{Charge-to-mass ratios for Au, C, and the difference between these two elements rescaled to our estimate for the SC composition.}
    \label{tab:charge}
\end{table}

Now let us piece together our detailed knowledge of the LPF sensitivity with our signal prediction. 
Table~\ref{tab:charge} shows the elemental charge-to-mass ratios for carbon and gold \cite{nist}. We ignore the Pt contribution to the TMs as its charge-to-mass ratio is close to the one of Au. 
We can derive the simple relation for the SC-TM difference of charge-to-mass ratio under the conservative assumptions about the SC composition of sec.~\ref{sec:LPF}:
\begin{align}
    \left(\frac{q}{M}\right)_{\text{TM}}=&\left(\frac{q}{M}\right)_{\text{Au}}\\
    \left(\frac{q}{M}\right)_{\text{SC}}\approx & \ f_\text{C}\left(\frac{q}{M}\right)_{\text{C}}+(1-f_\text{C})\left(\frac{q}{M}\right)_\text{Au}\\
    \left|\Delta \left(\frac{q}{M}\right)\right|=&\left| \left(\frac{q}{M}\right)_{\text{TM}}-\left(\frac{q}{M}\right)_\text{SC}\right|
    \approx f_\text{C}\left| \left(\frac{q}{M}\right)_\text{Au}-\left(\frac{q}{M}\right)_\text{C}\right|\;,
\end{align}
with $f_\text{C}\approx 83\,\text{kg}/450\,\text{kg}\approx 0.18$.
The last column of Table~\ref{tab:charge} shows the corresponding absolute value of this difference.\footnote{It is an unfortunate coincidence that the baryon charge-to-mass ratio is so similar for Au and C. Taking into account the true composition of the SC will alleviate this suppression.} 
Finally, we observe that our initial estimate of the suppression in the $B$ charge-to-mass ratio is in good agreement with the actual calculation.

Demanding that the SNR in eq.~\eqref{eq:snr} is at most unity we get a good estimate for the LPF sensitivity on the coupling strength of the DP to the chosen gauge group. For the DM density, we assume $\rho_\text{DM}\simeq 0.4\,$GeV/cm$^3$ \cite{deSalas:2019pee}.

As noted earlier, there is a rigorous way to combine the different axes but it requires taking into account a proper convolution of SC position and orientation with all possible DP polarizations. Whereas this procedure necessitates knowledge of the exact orbit of LPF it will provide even stronger limits if one follows the detailed guide provided in Ref.~\cite{Caputo:2021eaa}. LPF offers the advantageous feature that it is sensitive in all three spatial dimensions which means that our results cannot suffer from a ``blindness'' due to an unfortunate orientation of the polarization. We can always set a conservative estimate from taking the least sensitive axis for every frequency according to fig.~\ref{fig:asd}.

\begin{figure}[t]
    \centering
    \includegraphics[scale=0.5]{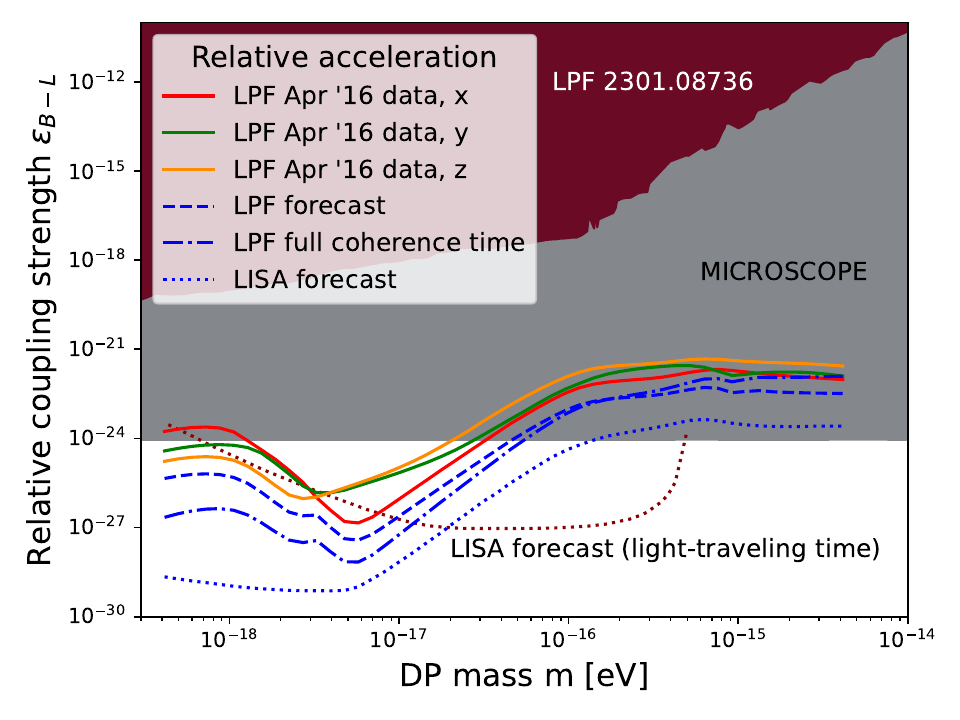}
    \caption{Limits on the rescaled coupling to $B-L$, $\epsilon_{B-L}$, of DPDM. In grey we show the DM-independent limits from searches for violation of the Equivalence Principle \cite{MICROSCOPE:2022doy} and the dark red filled region shows the LPF limits derived from decoherence in Ref.~\cite{Miller:2023kkd}. The dark red dotted line shows the forecast from LISA \cite{Morisaki:2020gui}. In red, green and orange we show the main result of this paper. Forecasts for similar analyses are shown in blue using the envelope simplification explained in the text.}
    \label{fig:b-l}
\end{figure}

In fig.~\ref{fig:b-l}, we show our main result for $B-L$ as solid lines for the individual axes following the color-coding of fig.~\ref{fig:asd}. Here we assume for each axis separately that the polarization is exactly aligned with the given axis, i.e.\ setting $\cos\theta_{A,i}=1$ in eq.~\eqref{eq:axacc}. We see that we get rather similar constraints from all axes except for the better peak sensitivity of the x-axis. Following the above argument by taking the upper envelope, i.e.\ to just consider the weakest limit for every mass, we can obtain a conservative combination of the limits. 

\begin{figure}[t]
    \centering
    \includegraphics[scale=0.5]{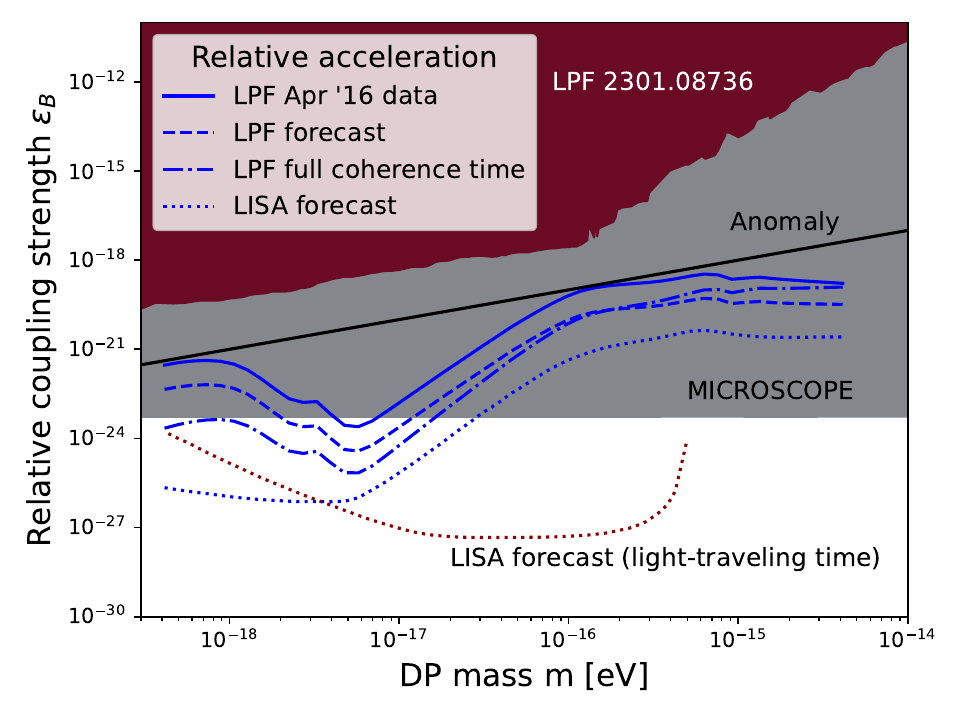}
    \caption{Limits on the rescaled coupling to $B$, $\epsilon_B$, of DPDM. While most of the limits in fig.~\ref{fig:b-l} are quite similar for baryon number, there are additional limits from the anomalous nature of this gauge group \cite{Dror:2017ehi} shown as a solid black line.}
    \label{fig:b}
\end{figure}

\begin{figure*}[t]
    \centering
    \includegraphics[scale=0.4]{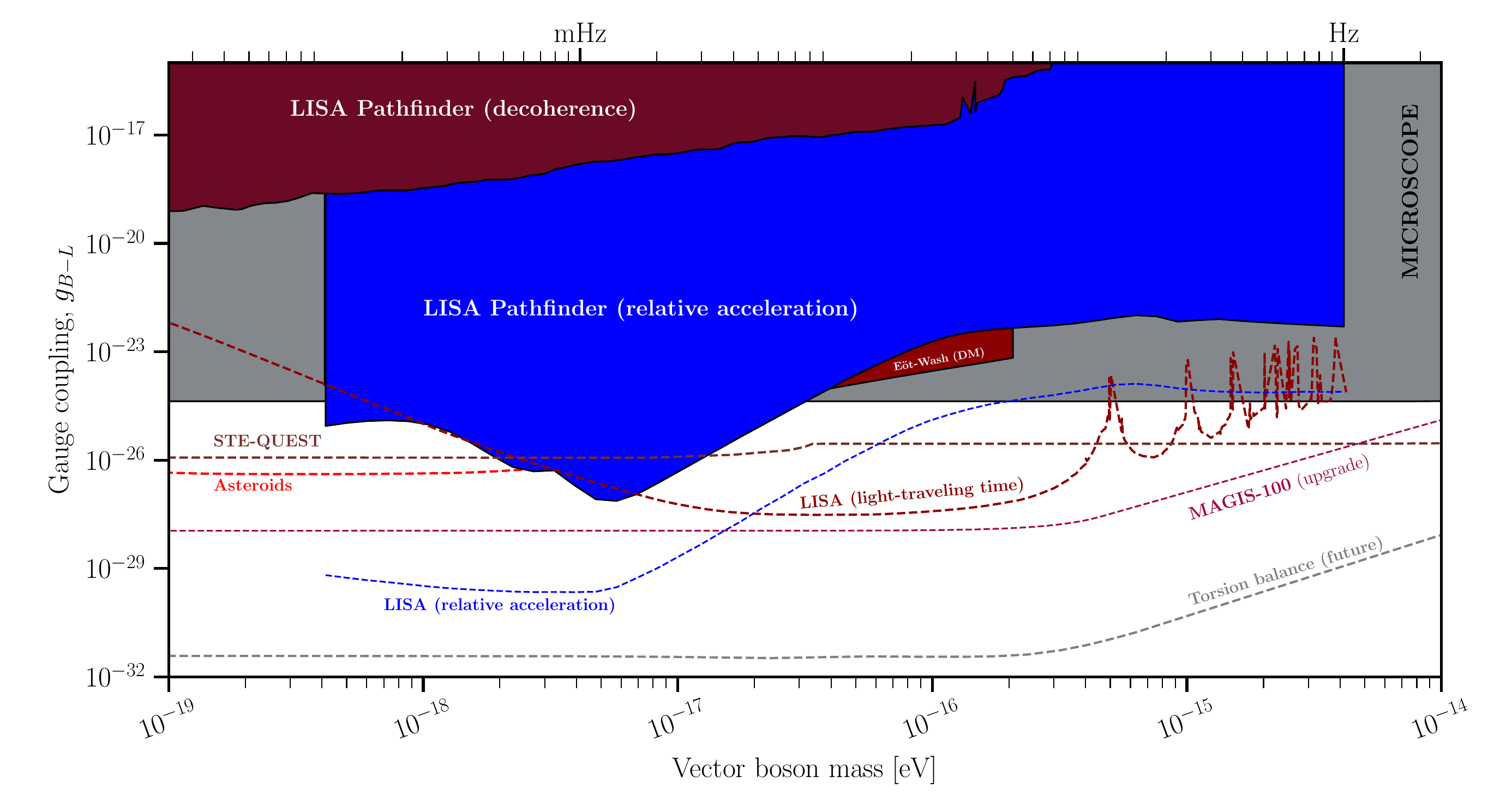}
    \caption{Limits on the gauge coupling of $B-L$ DPDM. These now include additional projections and limits as presented in \cite{AxionLimits}.}
    \label{fig:ohare}
\end{figure*}

Keeping this in mind, we will nevertheless opt for a more optimistic way to simplify the visualization of additional forecasts and later the results for $B$. Following Ref.~\cite{Pierce:2018xmy} we perform an average over all possible velocities and polarizations. While this is technically not the most conservative assumption for these long coherence times, we adopt this approach to facilitate comparison with previous LPF limits \cite{Miller:2023kkd} and LISA projections \cite{Pierce:2018xmy,Morisaki:2020gui} (see also appendix A of Ref.~\cite{Fedderke:2022ptm} or Ref.~\cite{Morisaki:2018htj}.). As our limits are independent of the velocity of the DPs, the resulting ``geometry factor'' is $1/\sqrt{3}$ as compared to the usual result of $1/3$. Then, instead of taking the upper envelope we will use the lower envelope, i.e.\ the strongest limit for every mass, multiplied by this suppression factor. We will refer to this as the envelope simplification.\footnote{Most of the limits we show with this method are more optimistic projections anyways. Only for the solid blue line of the $B$ limits in fig.~\ref{fig:b} and the blue region in fig.~\ref{fig:ohare} we should keep the shape of all three axes in mind.}

Using this approach we also include an estimate of the improved reach of LPF as a dashed blue line taking into account the whole data set and using the improved understanding of the detector noise and reduction of Brownian noise in the later stages of the mission \cite{Armano:2018kix}.\footnote{For this we set the observation time to 1 month and assume an improvement in noise-reduction by a factor 3.} We demonstrate the impact of the observation time by also including the blue dotted line which assumes the same sensitivity as the solid lines but we set the observation time to the coherence time for each frequency. This explains why the high frequency sensitivity is similar to the "fixed observation time" scenario: for the highest frequencies available, i.e.\ around $1\,$Hz, the coherence time is about $10^6\,$s which is roughly on the time scale of a week, coinciding with the real observation time used to model the sensitivity curve. On the low frequency end of the spectrum, the coherence times approach millennia scales making the dashed line much stronger than the other limits.

To demonstrate the power of our approach we compare it to three major results from the literature. We see that our analysis is able to cover new parameter space beyond the otherwise dominant limits set by the fifth-force search interpretation of the MICROSCOPE experiment \cite{MICROSCOPE:2022doy}. Furthermore, it is immediately clear that our analysis can easily outperform previous LPF limits \cite{Miller:2023kkd} just because there is no need to rely on the decoherence of the field which is the dominant effect if one only considers the two test masses. In fact, the improvement of our limits over the naive results that can be obtained from the decoherence method evaluated around our peak sensitivity at $\sim 5\cdot 10^{-18}\,$eV is given by
\begin{align}
    \frac{\epsilon_{B-L,\text{sat}}}{\epsilon_{B-L,\text{dec}}}\sim \frac{\Delta \frac{q}{m}}{\frac{q}{M}}\left(mvL\right)^{-1}\sim 3\cdot 10^{12}\;,\label{eq:boost}
\end{align}
ignoring the small difference in sensitivity between the $x_{12}$ and the $x_1$ sensitivity at this mass. The first factor takes into account that our method suffers from a mild charge-to-mass ratio suppression w.r.t.\ the decoherence method whereas the second factor comes from smallness of the decoherence on a length scale of $40\,$cm. We note that the limits found in Ref.~\cite{Miller:2023kkd} are better than naively expected from our analysis method, presumably because of their more sophisticated statistical analysis. This observation makes us confident in the potential reach of our approach for future analyses using all the available data. The third literature result is a LISA forecast using the conventional analysis method \cite{Morisaki:2020gui}.

Before turning to the LISA projections let us discuss the results for $B$ shown in fig.~\ref{fig:b}. We lose around 3 orders of magnitude in sensitivity which can be explained by the stronger charge-to-mass ratio suppression in eq.~\eqref{eq:boost} for $B$. This becomes immediately clear in the comparison of our results to the decoherence limits from LPF which do not suffer from this issue as they scale with the total charge-to-mass ratio. Nevertheless, the decrease in sensitivity of the Equivalence Principle limits due to the same effect still allows to probe a small region of new parameter space and makes an extended study of LPF (and LISA) auxiliary channels very attractive as it will cover a significant amount of new parameter space. Additionally, we added limits from the baryon number anomaly \cite{Dror:2017ehi} which are non-existent for $B-L$.

The limits shown as solid lines are rather robust and include conservative estimates on several different levels. Now we will take a more optimistic point of view and focus especially on the future LISA mission. As noted earlier in eqs.~\eqref{eq:snr} \& \eqref{eq:teff}, longer observation times up to one coherence time are extremely efficient to enhance the limits. With the peak sensitivity of LPF lying at around $10^{-3}\,$Hz it would be ideal to have data for around 30 years which of course is far beyond the actual lifetime of the mission. Nevertheless, the LISA mission may take data for up to 10 years \cite{LISA:2017pwj} which means that it naively maximizes the efficiency for frequencies around $\sim3\cdot 10^{-3}\,$Hz. Together with a general decrease in the noise this will allow for probing the ultra-low frequency parameter space complementary to the previous LISA forecasts.\footnote{Optimistically, we will assume a factor 10 improvement from the LPF sensitivity in 2016 and mitigation of the star tracker noise for our projections.} 

Previous projections using the planned arm length of around $2.5\cdot 10^6\,$km significantly cut into unexplored parameter space as shown in figs.~\ref{fig:b-l} \& \ref{fig:b}. These limits are based on looking for TM-TM displacements using the light-traveling time method and they are strongest around masses of $10^{-16}\,$eV. However, decreasing the mass by just one order of magnitude already introduces a decline of the limits by a factor of at least 100. In contrast to that, our method is well-suited for the lowest frequencies available because eq.~\eqref{eq:snr} does not depend on the arm length at all. Thus, there is no suppression of the constraints for low frequencies, i.e.\ large coherence lengths, except for the intrinsic sensitivity loss of the instrument. Indeed, the enhanced reach of our limits at small masses agrees very well with the findings of Ref.~\cite{Michimura:2020vxn} using the KAGRA auxiliary channels. Even though these auxiliary channels are at best as sensitive as the main interferometer they clearly outperform the conventional limits in the low mass region. In conclusion, the LISA mission will provide us with a very powerful tool to constrain the interactions of DPDM when combining the main channel analysis with the auxiliary channel analysis. These limits, spanning several orders of magnitude in mass, will reach deeply into unprobed parameter space.

\bigskip

As mentioned in sec.~\ref{sec:signal} our approach is not limited to $B-L$ and $B$. In fact, there is a plethora of additional gauge groups $g$ that will have very similar limits. These limits just require a proper rescaling procedure depending on the type of coupling. As noted earlier, there are essentially two types of couplings in our problem when it comes to analyzing observations involving different elements. The first one (``$B-L$-like'') is essentially sensitive to different neutron-to-proton ratios of the different elements while the second one (``$B$-like'') relies on the smaller differences in binding energies for different nuclei. Limits that instead depend on the total charge-to-mass ratio do not suffer from this ``binding energy suppression'' as can be seen from the small changes between the LISA projections and the previous LPF limits in fig.~\ref{fig:b-l} to fig.~\ref{fig:b}. For arbitrary gauge groups with a given combination of baryon and lepton number $\alpha B-\beta L$ we find that the charge-to-mass ratio can change from element to element.\footnote{For simplicity, we take $L=L_e$ here as $L_\mu$ and $L_\tau$ will give no contribution.} Therefore, ignoring different isotopes and changing to nuclear physics notation we find for an element with atomic number $Z=L$ and mass number $A=B$
\begin{align}
    \left(\frac{q}{M}\right)_{\alpha B - \beta L}\simeq &\frac{\alpha A -\beta Z}{ A m_p}=\alpha\frac{1}{m_p}-\beta\frac{Z/A}{m_p} \;,
\end{align}
where $m_p$ denotes the proton mass.

If instead we are interested in the difference between two elements we find
\begin{align}
    \Delta \left(\frac{q}{M}\right)_{\alpha B - \beta L}= \begin{cases}
        \alpha \Delta\left(\frac{q}{M}\right)_B\ \ ,& \beta=0\\
        \beta \Delta\left(\frac{q}{M}\right)_{B-L}\ \ ,& \text{else}
    \end{cases}\;,
\end{align}
using our results for the total charge-to-mass ratios from before. This makes the distinction between ``$B-L$-like'' and ``$B$-like'' immediately clear. Only a gauge group without coupling to electron number will suffer from the binding energy suppression. The interesting observation is that the calculation for $\beta\neq 0$ is already enough to rescale our limits to all possible gauge groups fulfilling this criterion. We present a selection of groups in table~\ref{tab:rescale}. The second column shows the rescaling for the relative charge-to-mass ratio and the third column shows the rescaling for the total charge-to-mass ratio. We note that the decoherence/light-traveling limit rescalings are technically only valid for Au and the $\beta=0$ are only valid for Au-C systems. Nevertheless, the rescalings for these cases will still give solid approximations for the true rescaling factor to arbitrary elements.

\begin{table*}[t]
    \centering
    \begin{tabular}{lc c c c}
    \hline\hline
         Limits & Type of coupling & Relative Acceleration & Decoherence/Light-traveling time (Au)\\ 
         \hline
       $L_e-L_\mu$ & $B-L$ &  1  &  $79/(197-79)\approx0.67$\\
       $L_e-L_\tau$ & $B-L$ & 1 &  $79/(197-79)\approx0.67$\\
       $B-3L_e$ & $B-L$ & 3 &$(3\cdot 79-197)/(197-79)\approx0.34$\\
       $B-3L_\mu$ & $B$ & 1 & 1 \\
       $B-3L_\tau$ & $B$ & 1 & 1 \\
       \hline\hline
    \end{tabular}
    \caption{Recipe to rescale the limits for more gauge groups. We give the type of coupling and the corresponding rescaling factor.}
    \label{tab:rescale}
\end{table*}

\bigskip

In sec.~\ref{sec:signal} we neglected any contribution from kinetic mixing. Let us briefly discuss the main reasons why this is well justified. First of all, in-medium effects lead to an effective suppression of the kinetic mixing if the plasma mass $\omega_p=\sqrt{4\pi \alpha n/m_e}$ is larger than the DP mass \cite{Redondo:2008aa}, i.e. $\epsilon_\text{KM,eff}\propto m^2/ \omega_p^2$. $m_e$ denotes the electron mass and $n\sim 5e^-/$cm$^3$ denotes the electron density in the interplanetary medium close to Earth \cite{ISSAUTIER20052141} implying a plasma mass of $\sim 10^{-10}\,$eV which is much larger than our mass range of interest. Secondly, both the TMs and the SC are essentially electrically neutral \cite{LISAPathfinder:2022ich}. Finally, the SC acts like a Faraday cage for the TMs \cite{Chaudhuri:2014dla}. Of course, the plasma will also interact with gauged DPs \cite{Heeba:2019jho} but if we consider the plasma mass of the DPs due to their direct coupling to SM particles $\omega_{p,g} \sim \epsilon_g \omega_{p}$ we see that the effects are very small.

Finally, let us put the constraints derived in this letter into larger context for gauged $B-L$ using the excellent collection of limits from Ref.~\cite{AxionLimits} shown in fig.~\ref{fig:ohare}. Note that the y-axis shows the gauge coupling $g_{B-L}=\epsilon_{B-L}e$. In addition to the limits shown above, one can also consider equivalence principle violation searches as direct detection experiments in a similar mass range \cite{Shaw:2021gnp}. These limits are quite similar to our work as they also search for a monochromatic DPDM signal on a ``$B-L$-dipole'' test mass. Several additional projections are shown in this plot coming from asteroids \cite{Fedderke:2022ptm}, atomic interferometry \cite{MAGIS-100:2021etm}, space-based quantum sensors \cite{STE-QUEST:2022eww}, and future torsion balance experiments \cite{Graham:2015ifn}. We see that neither LPF nor LISA is expected to have the best sensitivity in the long run but as LPF already has available data, this makes it the leading limit over almost two orders of magnitude in mass and at peak sensitivity it outperforms the current limits by more than two orders of magnitude in the gauge coupling. Furthermore, we have outlined why and how a detailed analysis of the LPF data can push the sensitivity providing excellent motivation for further work.

\section{Conclusion}\label{sec:con}

In this work we demonstrated how to improve existing DP limits based on the LPF data. The novel idea is that there is the option to use auxiliary measurements for the acceleration between the SC and the TMs to constrain the coupling strength of gauged $B-L$ and $B$ DPDM in analogy to the use of auxiliary arms in KAGRA \cite{Morisaki:2020gui}. The main advantage is the different atomic compositions of the test masses and the space craft leading to a relative acceleration. Relying only on the measurement between the two TMs will lead to extremely suppressed limits as the TMs react identically to the DPDM field. The existing literature focused on decoherence and light-traveling time effects which weakly break this degeneracy at the cost of a massive suppression for low frequencies where the arm length is much smaller than the wavelength. While our new limits are also moderately suppressed by the similar charge-to-mass ratios they are free from any arm length suppression and can therefore rely on the auxiliary channels working at almost full sensitivity. For LPF the auxiliary channels at their peak frequency are not significantly more noisy than the main ($x_{12}$) channel which is an important advantage for our work. Furthermore, we can cover all three spatial dimension with the auxiliary channels which will prevent a potential blindness towards specific DP polarizations.

We showed that even conservative estimates of the LPF results are already able to probe much new parameter space in the $B-L$ case and at least a small region for $B$ considering masses around $5\cdot10^{-18}\,$eV. Our approach offers an enhancement of up to about 12 orders of magnitude over the most naive analysis of $B-L$. It is therefore likely that a detailed analysis of the whole data set of LPF will set even better and thus world-leading limits over a considerable mass range. Additionally, this work motivates a rigorous analysis of the reach of LISA using auxiliary channels as our approach might be highly complementary to the previous forecasts.

\section*{Acknowledgements}
We would like to thank Andreas Ringwald for useful discussions. JF is grateful to Fermilab for its hospitality. This project has received funding from the European Union’s Horizon Europe research and innovation programme under
the Marie Skłodowska-Curie Staff Exchange grant agreement No 101086085 -- ASYMMETRY and the ITN HIDDeN grant agreement No 860881 -- HIDDeN, and from the Deutsche Forschungsgemeinschaft (DFG, German Research Foundation) under Germany’s Excellence Strategy -- EXC 2121 “Quantum Universe” -- 390833306 and through the Emmy Noether Grant No. KA 4662/1-2.

\bibliographystyle{elsarticle-num} 
\bibliography{bib}

\end{document}